\documentclass[prd,nofootinbib,preprint,floatfix]{revtex4}
\usepackage{amsmath,amssymb}
\usepackage[normalem]{ulem}
\usepackage{graphicx}
\usepackage{array}
\usepackage{color}

\newcolumntype{C}{>{~$}c<{$~}}
\newcolumntype{R}{>{~$}r<{$~}}

\def\lsim{\raise0.3ex\hbox{$\;<$\kern-0.75em\raise-1.1ex
\hbox{$\sim\;$}}}
\def\gsim{\raise0.3ex\hbox{$\;>$\kern-0.75em\raise-1.1ex
\hbox{$\sim\;$}}}
\DeclareMathAlphabet{\mathsc}{OT1}{cmr}{m}{sc}

\begin{document}
\preprint{YITP-SB-08-08}

\vspace*{.25in}
\title{Constraints from Solar and Reactor Neutrinos on Unparticle Long-Range 
Forces}

\author{M.C.~Gonzalez-Garcia}
\email{concha@insti.physics.sunysb.edu}
\affiliation{
C.N.~Yang Institute for Theoretical Physics, 
SUNY at Stony Brook, Stony Brook, NY 11794-3840, USA}
\affiliation{
Instituci\'o Catalana de Recerca i Estudis Avan\c{c}ats (ICREA), 
Departament d'Estructura i Constituents de la Mat\`eria,
Universitat de Barcelona, 647 Diagonal, E-08028 Barcelona, Spain}
\author{P.C.~de Holanda} 
\email{holanda@ifi.unicamp.br}
\affiliation{
Instituto de F\'{\i}sica Gleb Wataghin,
Universidade Estadual de Campinas,
C.\ P.\ 6165, 13083-970, Campinas, Brazil}
\author{R. Zukanovich Funchal}
\email{zukanov@if.usp.br}
\affiliation{
Instituto de F\'{\i}sica, Universidade de S\~ao Paulo, 
 C.\ P.\ 66.318, 05315-970 S\~ao Paulo, Brazil}
\begin{abstract}
  We have investigated the impact of long-range forces induced by
  unparticle operators of scalar, vector and tensor nature coupled to
  fermions in the interpretation of solar neutrinos and KamLAND
  data. If the unparticle couplings to the 
  neutrinos are mildly non-universal, such long-range forces will not
  factorize out in the neutrino flavour evolution. As a consequence
  large deviations from the observed  standard matter-induced oscillation
  pattern for solar neutrinos would be generated. In this case, 
  severe limits can be set on the infrared fix point scale, $\Lambda_u$, 
  and the new physics  scale, $M$, as 
  a function of the ultraviolet ($d_{\rm UV}$) and anomalous ($d$) dimension 
  of the unparticle operator. For a scalar unparticle, for instance, 
  assuming the non-universality of the lepton couplings to unparticles 
  to be of the order of a few per mil  we find  that, for $d_{\rm UV}=3$ and 
  $d=1.1$,  $M$ is constrained to be
  $M> {\cal O}(10^{9})$ TeV ($M> {\cal O}(10^{10})$ TeV) 
  if $\Lambda_u =$ 1 TeV (10 TeV).
  For given
  values of $\Lambda_u$ and $d$, the corresponding bounds on $M$
for  vector (tensor) unparticles  are 
  $\sim 100$ ($\sim 3/\sqrt{\Lambda_u/{\rm TeV}}$) times those
for the scalar case.
  Conversely, these results can be translated into severe constraints 
  on universality violation of the fermion couplings to unparticle 
  operators with scales which can be accessible at future colliders. 
\end{abstract}
\maketitle
\section{Introduction}
\label{sec:intro}

We are accustomed to describe fundamental interactions in nature in
terms of quantum fields which manifest themselves as particles.  Our
most successful description of all the available experimental data
today is given by the Standard Model (SM) which is based on such
quantum fields. Extensions of the Standard Model, such as supersymmetric, 
grand-unified, or extradimensional models,  also rely on these
ideas and give rise to a plethora of new particles that may be
discovered soon at the LHC. However, other types of quantum fields may
be present in Nature.

Recently, Georgi~\cite{georgi1,georgi2} proposed a novel scenario
where no new particle states emerge but where scale invariant fields
with non-trivial infrared (IR) fixed point can be part of physics
above the TeV scale. If these fields couple to SM operators they can
give rise to effective interactions at lower energies that may have
impact on particle phenomenology.  He denominated this scale invariant
matter coupled to SM fields at lower energies {\em unparticle}, as it is 
not constrained by a dispersion relation like normal particles.

Since then there has been a number of papers in the literature
devoted to map the unusual observable consequences of unparticles to
experiments. QED bounds to unparticle interactions have been obtained
in \cite{unpar:qed}. Signals of unparticle have been studied in
past and future collider
experiments~\cite{cheung,unpar:collider}.  Possible
unparticle contributions to CP violation~\cite{cpviol}, deep inelastic
scattering~\cite{dis}, lepton flavour violating processes~\cite{lfv} as
well as to hadron mixing and decay~\cite{hadron}, neutrino
interactions~\cite{neutrinos,kam} and nucleon decay ~\cite{pakvasa}
 have been investigated. Several
cosmological and astrophysical bounds have also been
devised~\cite{cosmo1,cosmo,wyler}.

Another striking spin-off of the unparticle idea is the advent of
long-range forces operating at macroscopic distances and generally
governed by a non-integral power law. This unique behavior was first
pointed out in Ref.\cite{spin}, in connection to microscopic
spin-dependent interactions between electrons and has also been
investigated in connection to deviations of the Newtonian
gravitational inverse squared law mediated by a tensor-like {\em
  ungravity} interaction~\cite{lr1}. Limits on vector-like unparticle
interactions coupled to baryon and lepton flavour numbers were
established in \cite{lr2} by comparing with results from
torsion-balance experiments. 

Neutrino oscillations have been shown to be a very powerful tool in
constraining new dynamics. Being an interference phenomenon it can be
sensitive to very feeble interactions as their effect enters linearly
in the neutrino propagation. Furthermore the effect of new fermionic
interactions can be enhanced when neutrino travels regions densely
populated by those fermions.  In this way it has been shown  that
neutrino oscillation data can place stringent bounds on long-range
forces coupled to lepton flavour
numbers~\cite{ourlrange,others,mohanty}.  Therefore it is expected
that they can provide very strong limits also on the couplings of
unparticles to fermionic currents.

Assuming that scale invariance is not broken in the IR we 
derive in this paper, from solar and reactor neutrino 
experimental data, limits on scalar, vector and tensor
unparticle operators which mediate long-range interactions 
through their couplings  to fermionic flavour-conserving currents.
In Sec.~\ref{sec:form} we briefly present the formalism of unparticle
dynamics relevant to the present study. We present the effects of
the induced long-range forces in solar neutrino oscillations in 
Sec.~\ref{sec:solarosc}. We derive the bounds from the no observation
of such long-range effects from the global analysis of solar and
KamLAND data in Sec.~\ref{sec:const} and in the final Section we discuss 
our conclusions.

\section{Unparticle Formalism in Brief}
\label{sec:form}

We will consider hypothetical fields of a hidden sector of a theory with
a non-trivial IR fixed point which can interact with SM fields at very
high energies by the exchange of a particle with a large mass $M$ as
proposed in Ref.~\cite{georgi1}. So the ultraviolet (UV) couplings of
this hidden sector operators ${\cal O}_{\rm UV}$ of dimension $d_{\rm
UV}$ with SM operators ${\cal O}_{\rm SM}$ of dimension $d_{\rm SM}$
take the form
\begin{equation}
\frac{1}{M^{d_{\rm UV}+d_{\rm SM}-4}} \, {\cal O}_{\rm UV} 
\, {\cal O}_{\rm SM}\, .
\end{equation}

Scale invariance emerges at a lower energy scale $\Lambda_u$. As
the hidden sector runs towards the IR fixed point, these couplings are
so to speak {\em sequestered}, their scaling dimension is increased by
the anomalous dimension $d$ from the hidden sector dynamics. From this
point on the hidden sector operator matches onto an unparticle
operator ${\cal O}_{\rm U}$ whose interactions with the SM operators
are described by the non-renormalizable effective lagrangian

\begin{equation}
 C_{u} \frac{\Lambda_{u}^{d_{\rm UV}-d}}{M^{d_{\rm UV}+d_{\rm SM}-4}} \, {\cal O}_{\rm SM} \, {\cal O}_{\rm U}\, ,
\end{equation}
where $C_{u}$ is a dimensionless constant of order 1.

Our starting point will be the lagrangian which contains the effective
flavour diagonal interactions which satisfy the SM gauge symmetry,
between scalar, vector and tensor unparticles and the SM fermions $f$
\begin{equation}
{\cal L}=-\sum_f\frac{\lambda^f_S}{\Lambda_u^{d-1}} \bar f f{\cal O}_U
-\sum_f \frac{\lambda^f_V}{\Lambda_u^{d-1}} \bar f\gamma_\mu f{\cal O}^\mu_U
-\sum_f i \frac{\lambda^f_T}{2 \Lambda_u^{d}} 
(\bar f\gamma_\mu \partial_\nu f- \partial_\nu \bar f \gamma^\mu
f){\cal O}^{\mu \nu}_U \, ,
\label{eq:lagrangian}
 \end{equation}
where the unparticle operators are taken to be hermitian. Additional 
four fermion contact interactions could also arise as discussed
in Ref.~\cite{grinstein}. In the absence of cancellations they will
add up to the effects discussed here making the unparticle effects
larger. In what follows we will conservatively neglect the
contributions from contact interactions. 

The natural size of the $\lambda^f_{S,V,T}$ coefficients is
\begin{equation}
\lambda^f_{S,V,T}=C^f_{S,T,V}\left(\frac{\Lambda_u}{M}\right)^{d_{UV}-1}\, ,
\end{equation}
where $C^f_{S,T,V}$ are again of order 1.

Scale invariance determines the propagation properties of these
operators up to a normalization factor. Using the same convention as
in Ref.~\cite{georgi2,cheung}, including the corrections given 
in \cite{grinstein}, 
we obtain that the propagators corresponding to
scalar, vector and tensor unparticle operators are correspondingly
\begin{eqnarray}
&&\Delta (P^2) 
=\frac{A_d}{2\sin(d \pi)} (-P^2)^{d-2} \, ,\\
&&[\Delta (P^2)]_{\mu\nu}=\Delta (P^2)\; \pi_{\mu\nu} \, , \\
\text{and} &&[\Delta (P^2)]_{\mu\nu,\rho\sigma}=\Delta (P^2)\; T_{\mu\nu,\rho\sigma} \, .\\
\end{eqnarray}
\begin{equation}
\pi_{\mu\nu}=-g_{\mu\nu}+ a 
\, \frac{P_\mu P_\nu}{P^2} \, ,
\end{equation}
where $a=1$ for transverse ${\cal O}^\mu_U$ and $a=\frac{2(d-2)}{d-1}$
in conformal field theories~\cite{grinstein}. 

For transverse and  traceless  ${\cal O}_U^{\mu \nu}$ operators 
\begin{equation}
T_{\mu\nu,\rho\sigma}=\frac{1}{2} \left\{
\pi_{\mu\rho}\pi_{\nu\sigma}+\pi_{\mu\sigma}\pi_{\nu\rho}-
\frac{2}{3}\pi_{\mu\nu}\pi_{\rho\sigma}\right\} \, ,
\end{equation}
while in conformal field theories
\begin{eqnarray}
T_{\mu\nu,\rho\sigma}&=&\frac{1}{2} 
\bigg[\left(g_{\mu\rho}g_{\nu\sigma}+\mu\leftrightarrow\nu\right)
+\frac{\left[4-d(d+1)\right]}{2d(d-1)}g_{\mu\nu}g_{\rho\sigma}
\\ \nonumber
&&
-2\frac{(d-2)}{d}\left(g_{\mu\rho}\frac{k_\nu k_\sigma}{k^2}
+g_{\mu\sigma}\frac{k_\nu k_\rho}{k^2}+\mu\leftrightarrow\nu\right)
\nonumber \\
&& +4\frac{(d-2)}{d (d-1)}\left(g_{\mu\nu}\frac{k_\rho k_\sigma}{k^2}
+g_{\rho\sigma}\frac{k_\mu k_\nu}{k^2}\right) 
+8\frac{(d-2)(d-3)}{d(d-1)}
\frac{k_\mu k_\nu  k_\rho k_\sigma}{(k^2)^2}\bigg]~ \;.
\end{eqnarray}
In the equations above 
\begin{equation}
A_d=\frac{16 \pi^2 \sqrt{\pi}}{(2\pi)^{2 d}}\frac{\Gamma(d+\frac{1}{2})}
{\Gamma(d-1)\Gamma(2d)}\, .
\end{equation}

These massless unparticle operators induce long-range forces between
fermions. In particular, the effective lagrangian describing the 
long-range force generated by a static distribution of fermions $f$
with number density $n_f(r)$, assumed to be spherically symmetric and 
of radius R, on a relativistic fermion $f'$ is given by

\begin{equation}
{\cal L}=\left[\frac{ C^f_S C^{f'}_S}{4 \pi}\bar f' f' -   
\frac{ C^f_V C^{f'}_V}{4 \pi}
\bar f'\gamma^0 f' +
  \frac{ C^f_T C^{f'}_T}{4 \pi} \frac{m_f E'}{\Lambda_u^2} \, B\, \bar f'\gamma^0 f' 
\right] W_f(r) \, ,
\label{eq:llr}
\end{equation}
where $B=2/3$ for a transverse and traceless unparticle tensor operator, 
while in conformal field theories $B=1-\frac{\left[4-d(d+1)\right]}{4d(d-1)}$.
The potential function, irrespectively of the Lorentz structure of the 
unparticle operator, can be shown to have a universal form as a function of 
the distance $r$ from the center of the matter distribution
\begin{eqnarray}
W_f(r=x R)&=&
\left(\frac{\Lambda_u}{M}\right)^{2(d_{UV}-1)} 
\frac{2}{\pi^{2(d-1)}}\frac{\Gamma[d+\frac{1}{2}]\Gamma[d-\frac{1}{2}]}
{\Gamma[2d]}
R^3  \frac{1}{\left(R \Lambda_u\right)^{2(d-1)}}
\nonumber \\ 
&&\times \frac{2}{r} \;  
\begin{array}{l}
\frac{1}{3-2d}\int_0^1 y \,n_f(y) 
\left[(x+y)^{3-2d}-|x-y|^{3-2d}\right] dy 
\;\;\;\;\;\; {\rm if}\; d\neq \frac{3}{2} \\[+0.5cm] \frac{1}{2}
\int_0^1 y \, n_f(y) 
\ln\left[\frac{(x+y)^2}{(x-y)^2}\right] 
\;\;\;\;\;\;\;\;\;\;\;\; \;\;\;\;\;\; \;\;\;\;\;\; 
\;\;\;\;\;\; \;\;\;\;\;\; \;\;\;\;\;\;  
{\rm if}\; d= \frac{3}{2}\, .
\label{eq:W}
\end{array}
\end{eqnarray}
For $d \to 1$ one recovers the Coulomb-type potential characteristic
of a massless particle exchange.  This potential function is always
finite for $d<2$ whereas for $d>2$ there are infinite contributions
that appear when $x=y$.  In order to render a finite answer for $d\geq
2$ this infinite contribution has to be canceled out by the
introduction of some counterterm. In principle, this could be done by
means of the effective four fermion contact interactions whose
couplings gets renormalized.  Technically this is equivalent to remove
the point $x=y$ from the integration in Eq.~(\ref{eq:W}). In principle
a finite contact contribution to the potential can remain. As
mentioned above in what follows we will neglect this contribution.
In practice we will only be quantitatively discussing the results for
$d<2$, since, as we will see below, the effects become very feeble for
larger dimensions. 

If the theory is not only scale invariant but also conformal invariant
the dimension $d$ of the unparticle operator is bounded 
from unitarity constraints~\cite{grinstein}. For scalar, vector and 
symmetric tensor operators $d>$ 1, 3 and 4, respectively. 
We will relax this constraint in our phenomenological analysis.

One must also notice that the Higgs coupling to scalar unparticles
generally breaks scale invariance at a scale close to the electroweak
scale ~\cite{fox,feng,Delgado:2008rq,Delgado:2007dx}.  Therefore for
the scalar unparticle effects to be relevant at solar and reactor
energies ($\sim$ MeV) discussed here, one must assume the Higgs-scalar
unparticle couplings to be suppressed. For vector and tensor
unparticles, the Higgs only couples to higher dimension operators
which relaxes the bound on the scale of scale-invariance breaking.

\section{Unparticle Effects in Neutrino Oscillations in the Sun}
\label{sec:solarosc}

Under the assumptions discussed in the previous section, unparticles
mediate a long-range force coupled to fermions in the way described by 
Eq.~(\ref{eq:llr}). Such force will affect neutrino oscillations in matter 
in a similar fashion to the leptonic forces discussed in 
Ref.~\cite{ourlrange}.  

In our description of the effect of unparticles in the evolution of
neutrinos in matter we have to take into account that conventional
mass-induced neutrino oscillations describe very well
atmospheric~\cite{atm}, K2K~\cite{k2k}, MINOS~\cite{MINOS},
solar~\cite{chlorine,sagegno,gallex,sk,sno,sno05} and
reactor~\cite{kamland,chooz} neutrino data as long as all three
neutrino flavours must participate of oscillations \cite{myreview}.
The mixing among neutrino mass eigenstates and flavour eigenstates is
encoded in the $3\times 3$ lepton mixing matrix that in the standard
parametrization takes the form
\begin{equation}
    U=\begin{pmatrix}
    1&0&0 \cr
    0& {c_{23}} & {s_{23}} \cr
    0& -{s_{23}}& {c_{23}}\cr\end{pmatrix}
    \begin{pmatrix}
    {c_{13}} & 0 & {s_{13}}e^{i {\delta}}\cr
    0&1&0\cr 
    -{ s_{13}}e^{-i {\delta}} & 0  & {c_{13}}\cr 
\end{pmatrix} 
\begin{pmatrix}
    c_{21} & {s_{12}}&0\cr
    -{s_{12}}& {c_{12}}&0\cr
    0&0&1\cr
\end{pmatrix}
\end{equation}
where $c_{ij} \equiv \cos\theta_{ij}$ and $s_{ij} \equiv
\sin\theta_{ij}$. 

The neutrino oscillation data is consistent with the following hierarchy of 
the neutrino mass squared  differences 
\begin{equation} \label{eq:deltahier}
    \Delta m^2_\odot = \Delta m^2_{21} \ll 
|\Delta m_{31}^2|\simeq|\Delta m_{32}^2|
=\Delta m^2_{\rm atm}.
\end{equation}
Recent KamLAND~\cite{kamland} and MINOS~\cite{MINOS} results lead to
$\Delta m^2_{21}/\vert \Delta m^2_{31}\vert \approx 0.03$ \cite{myreview}, 
so for 
KamLAND and solar neutrinos, the atmospheric neutrino oscillation scale 
contributions are completely averaged and the interpretation of 
these data in the three neutrino oscillation framework depends basically  
on $\Delta m^2_{21}$, $\theta_{12}$ and $\theta_{13}$. In contrast,  
atmospheric and long baseline neutrino oscillation experiments are 
driven mostly by $\Delta m^2_{31}$, $\theta_{23}$ and
$\theta_{13}$.  Moreover, the non observation of oscillations in 
the CHOOZ  reactor experiment~\cite{chooz} imply that $\theta_{13}$, the 
mixing angle connecting solar and atmospheric neutrino oscillation 
channels must be very small, {\it i.e.}, $\sin^2\theta_{13} \leq 0.041$ 
at 3$\sigma$~\cite{myreview}.  These considerations make that 
the 3-$\nu$ oscillations effectively factorize into 2-$\nu$ oscillations 
of the two different subsystems: solar plus reactor, and atmospheric 
plus long baseline. 

Because the new couplings are flavour diagonal, 
their effects in the evolution of atmospheric neutrinos do 
not change the hierarchy (\ref{eq:deltahier}) ~\cite{mohanty}. Thus  
the 2$\nu$ oscillation  factorization still holds and as long
as we neglect the small $\theta_{13}$ we can 
perform the analysis of solar and reactor neutrinos in an effective
two-neutrino mixing scenario.   
In this framework, one can easily read from Eq.~\eqref{eq:llr} 
the effect  of the long-range unparticle force in the evolution of  
solar neutrinos as: 
\begin{eqnarray} 
i \frac{d}{dr} 
\left(\begin{array}{c} \nu_e \\
\nu_a     \end{array}\right)=&& 
\left\{ 
\frac{1}{2E_\nu}
\left[
     \mathbf{U}_{\theta_{12}}
\left(    \begin{array}{cc}
        m_1 & ~0 \\
        ~0& m_2 
    \end{array}\right)
\; 
     \mathbf{U}_{\theta_{12}}^\dagger 
-\left(    \begin{array}{cc}
        M_{\rm UNP}(r) & ~0 \\
        \hphantom{-} 0 & ~0 
    \end{array}\right)\right]^2 \right.    
\nonumber \\
&&\left.
+\left( \begin{array}{cc}
        V_{\rm CC}(r)+ V_{\rm UNP}(r) & ~0 \\
        \hphantom{-} 0 & ~0 
    \end{array}\right)\right\}
\left(\begin{array}{c} \nu_e \\
\nu_a    \end{array}\right),
\label{eq:evol}
\end{eqnarray}
where  $E_\nu$ is the neutrino energy, 
$m_{1,2}$ are the neutrino masses in vacuum, 
and $\mathbf{U}_{\theta_{12}}$ the mixing 
matrix between neutrino flavour and the vacuum mass eigenstates and
$\nu_a=c_{23}\nu_\mu +s_{23} \nu_\tau$. 
$V_{\text{CC}}(r)=\sqrt{2} G_F n_{e}(r)$ is the 
Mikheyev-Smirnov-Wolfenstein (MSW) matter potential.  

In Eq.~\eqref{eq:evol} 
$M_{\rm UNP}$ and $V_{\rm UNP}(r)$ depend on the Lorentz structure 
of the unparticle operator. 
The function $W$ will enter either  as an extra mass term 
(scalar unparticle) or as an addition 
to the MSW potential (vector or tensor unparticle).
For scalar unparticles,
\begin{eqnarray}
&& M_{\rm UNP}\equiv M_\text{S}(r)=  
\frac{C^e_S (C^{\nu_e}_S-C^{\nu_a}_S)}
{4 \pi} W(r) \simeq  \alpha_S  \, W(r)\; ,  \;\;\;\;\;\;\;
V_{\rm UNP}(r)=0 \; ,
\end{eqnarray}
while for vector unparticles, 
\begin{eqnarray}
&&V_{\rm UNP}(r)\equiv  V_\text{V}(r)=\frac{C^e_V (C^{\nu_e}_V-C^{\nu_a}_V)}
{4 \pi} W(r) \simeq  \alpha_V   \, W(r)\; , 
\;\;\;\;\;\;\; M_{\rm UNP}=0 \; ,  
 \end{eqnarray}
and for tensor  unparticle operators, 
\begin{eqnarray}
&&V_{\rm UNP}(r)\equiv -
V_\text{T}(r)=-\frac{C^e_T (C^{\nu_e}_T-C^{\nu_a }_T)}
{4 \pi} \frac{m_e E_\nu}{\Lambda_u^2}\, B\, W(r) 
\simeq  -\alpha_T   \frac{m_e E_\nu}{\Lambda_u^2} \, W(r)\; ,  
\;\; M_{\rm UNP}=0  \;,
\end{eqnarray}
where $C^{\nu_a}_{S,V,T}=
c^2_{23} C^{\nu_\mu}_{S,V,T}+s^2_{23}C^{\nu_\tau}_{S,V,T}$.
In all cases,  $W(r)$ is given in Eq.~(\ref{eq:W}) with 
\begin{equation}
n_f(r)=n_e(r)
\Big[1+\frac{C^p_{S,V,T}}{C^e_{S,V,T}} R_p+
\frac{C^n_{S,V,T}}{C^e_{S,V,T}}\frac{Y_n(r)}{Y_e(r)}
R_n\Big]\,,
\end{equation}
with $R_{p,n}=1$ for scalar and vector unparticles and 
$R_{p,n}=m_{p,n}/m_e$ for tensor unparticles.
$Y_f(r)$ is the relative number density of fermion $f$, it is different
for neutrons than for protons and electrons because of the
change in composition along the Sun radius. Because of this 
composition change, not only the normalization 
but also the $W(r)$ profile (in $r$) has some dependence 
on the unparticle  couplings.  We have neglected this small unparticle 
effect in the profile of the potential and absorbed it in the definition of
effective $r$ average couplings: 
\begin{equation}
\alpha_{S,V,T}  = 
\frac{C^e_{S,V,T} (C^{\nu_e}_{S,V,T}-C^{\nu_a}_{S,V,T})}{4\pi}  
\Big[1+\frac{C^p_{S,V,T}}{C^e_{S,V,T}} R_p+
\frac{C^n_{S,V,T}}{C^e_{S,V,T}}
\langle\frac{Y_n(r)}{Y_e(r)}\rangle R_n\Big]\,,
\label{eq:alcoup}
\end{equation}
where $\langle\frac{Y_n(r)}{Y_e(r)}\rangle$ stands for the average of the
relative number densities along the neutrino trajectory.  

From Eq.~\eqref{eq:alcoup} we see that, as expected, the unparticle
effect on the solar neutrino evolution does not factorize out as long
as the unparticle coupling to neutrinos are non-universal. As a matter
of fact, under the same assumption, solar neutrinos may also invisibly
decay in their way from the Sun to the Earth as described in
Ref.~\cite{kam}. In what follows we will neglect the decay and
concentrate on the effects due to the long-range forces, which as we
will see, turn out to be more constraining.

\begin{figure}[t]
\includegraphics[width=4.in]{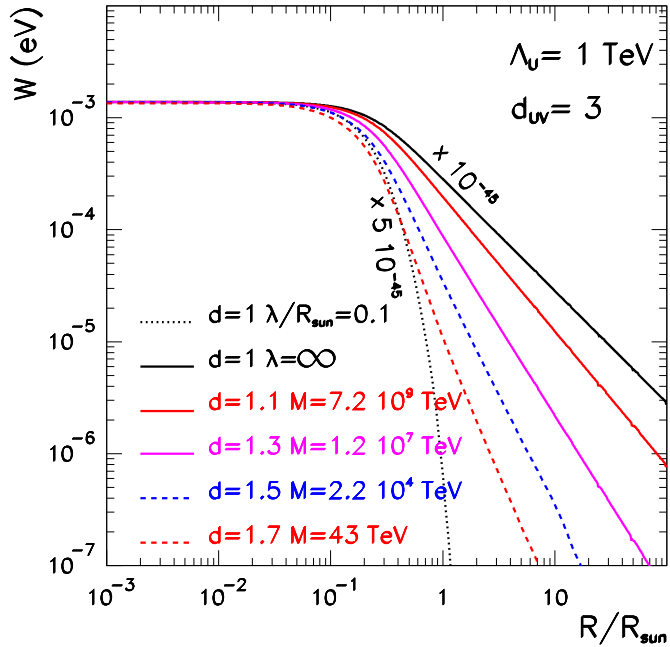}
\caption{{\sl Potential} function $W(r)$ due to the density of
  protons (or electrons) in the Sun as a function of the distance from the solar
  center in units of the solar radius, $R_{\odot}$, for various values
  of the dimension $d$ and the mass scale $M$. We take $\Lambda_u= 1$
  TeV and $d_{UV}=3$.  For comparison, we show the value of the $W(r)$
  function for a Coulomb-type ($d=1$) long-range force with range
  $\lambda=0.1 \, R_\odot$ and with the same strength at the center of
  the Sun.}
\label{fig:pot}
\end{figure}

In Fig.~\ref{fig:pot} we show the function $W(r)$ in the Sun as
function of the distance from the center in units of $R_{\odot}$.
As seen in the figure, the dimension $d$ acts as a kind of range of the
potential.  Inside the Sun and up to 0.1 $R_\odot$, $W$ is constant,
independently of $d$.  We notice that for any value of $d<1.7$, there is 
always a value of $M$ for which, at any $r$, the
$W(r)$ function is always larger than the corresponding one for a $d=1$
long-range force of range $\lambda=0.1$.

\begin{figure}[hbt]
\includegraphics[width=4.in]{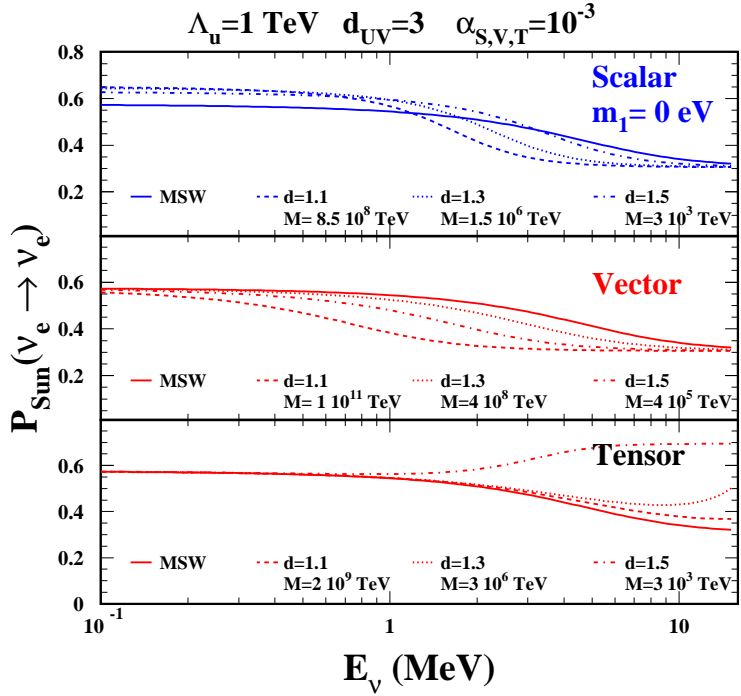}
\vglue -1.5cm
\caption{Survival probability of $\nu_e$ in the Sun as a 
function of the neutrino energy $E_\nu$ for a 
scalar (upper panel), vector (middle panel) and 
tensor (lower panel) unparticle force, for various 
values of the parameters. For all curves we have used 
$\tan^2\theta_{12}=0.44$ and $\Delta m^2_{21} =
7.9\times10^{-5}~{\text{eV}}^2$.} 
\label{fig:probs}
\end{figure}

We show in Fig.~\ref{fig:probs} the survival probability of solar 
$\nu_e$ at the sunny face of the Earth as a function of the neutrino 
energy $E_\nu$ for some values of the parameters. 
The probability is obtained by numerically solving Eq.~(\ref{eq:evol}) 
along the
neutrino trajectory from its production point in the Sun to its detection
point on the Earth. We have verified that for the range of parameters of
interest the evolution in the Sun and  from the Sun to the Earth is 
always adiabatic. From the figure we can infer the expected order
of magnitude of the bounds that can be derived from the 
solar and KamLAND analysis which we describe next.
For the scalar case we have chosen $m_1=0$ so  $m_2=\sqrt{\Delta m^2_{21}}$.
As shown in Ref.~\cite{ourlrange}, the effect of scalar interactions
becomes stronger as $m_1$ grows, so we chose  $m_1=0$ which gives
the most conservative constraints. 

\section{Constraints from Solar and Reactor Neutrino Data}
\label{sec:const}
We present in this section the results of the global analysis of solar
and KamLAND data for the unparticle long-range forces discussed in the
previous section.  Details of our solar neutrino analysis have been
described in previous papers~\cite{pedrosolar,myreview}.  We use the
solar fluxes from Bahcall, Serenelli and Basu 2005~\cite{BS05}.  The
solar neutrino data includes a total of 118 data points: the
Gallium~\cite{sagegno,gallex} and Chlorine~\cite{chlorine} (1 data
point each) radiochemical rates, the Super-Kamiokande~\cite{sk} zenith
spectrum (44 bins), and SNO data reported for phase 1 and phase 2.
The SNO data used consists of the total day-night spectrum measured in
the pure D$_2$O (SNO-I) phase (34 data points)~\cite{sno}, plus the
full data set corresponding to the Salt Phase (SNO-II)~\cite{sno05}.
This last one includes the neutral current and elastic scattering
event rates during the day and during the night (4 data points), and
the charged current day-night spectral data (34 data points). It is
done by a $\chi^2$ analysis using the experimental systematic and
statistical uncertainties and their correlations presented
in~\cite{sno05}, together with the theoretical uncertainties. In
combining with the SNO-I data, only the theoretical uncertainties are
assumed to be correlated between the two phases.  The experimental
systematics errors are considered to be uncorrelated between both
phases.

In the analysis of KamLAND, we neglect the effect of the long-range
forces due to the small Earth-crust density in the evolution
of the reactor antineutrinos. For KamLAND we have used the observed
events as a function of $L_0/E_\nu$, with $L_0$ fixed at 180 km 
~\cite{kamlandlast} and minimized the $\chi^2$ function

\begin{equation}
\chi^2_{\rm KL} = \sum_{i=1}^{24} \left[2 (f R^{\rm th}_{i} - R^{\rm ex}_i +
R^{\rm ex}_i \, \log \left( \frac{R^{\rm ex}_i}{f \, R^{\rm th}_i} \right) 
\right] + \left( \frac{1-f}{0.041}\right)^2
\,.
\label{eq:kam}
\end{equation}
with respect to $\Delta m^2_{21}$, $\tan^2 \theta_{12}$ and $f$. Here 
$R^{\rm ex}_i$ and $R^{\rm th}_i$ are, respectively, the observed and  
theoretical (which depends of the oscillation parameters) number of events, 
for the i-th bin.

One must also notice that the lagrangian given in
Eq.~(\ref{eq:lagrangian}) also leads to extra neutral current
contributions to neutrino interaction cross-sections at the detector.
However, we will ignore those effects which in the absence of
cancellations will add-up to the propagation effect.  In this sense the
bounds established here are conservative.

\begin{figure}[t]
\includegraphics[width=4.in]{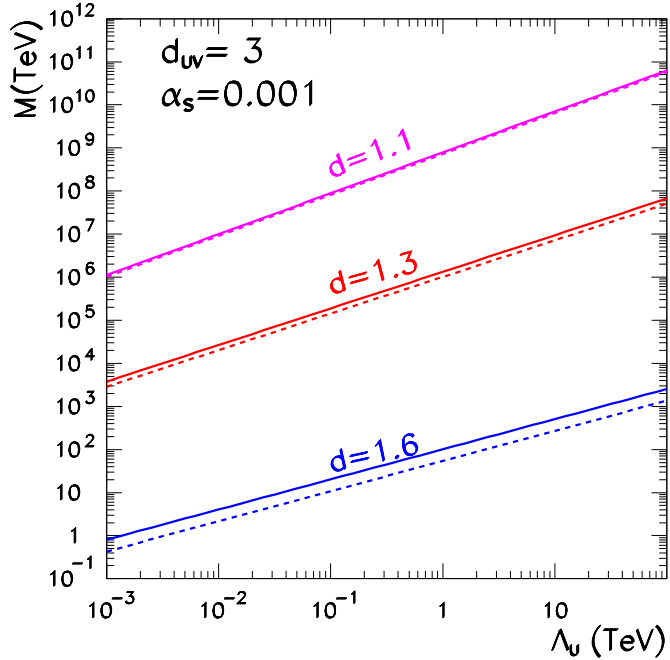}
\caption{Bounds from the analysis of solar and KamLAND data on the
fundamental parameters $(\Lambda_u,M)$ for a scalar unparticle for
$d=1.1,1.3,1.6$ and $d_{UV}=3$. The region below the curves is
excluded. We have fixed $\alpha_S=10^{-3}$. The dashed lines 
correspond to the approximation given in Eq.~\eqref{eq:ksd}.}
\label{fig:scalbounds}
\end{figure}
As an illustration we show in Fig.~\ref{fig:scalbounds} the bounds 
obtained from our combined analysis of solar and KamLAND data on the 
fundamental parameters for an scalar unparticle. For the sake of
concretness we have chosen $\alpha_{S}>0$ but similar order of
magnitude bounds can be derived for negative $\alpha_S$.

First we notice that the bounds from the long-range force effects 
are orders of magnitude stronger than the corresponding ones from the 
unparticle-induced invisible solar neutrino decay described in 
Ref.~\cite{kam}. This justifies our approximation of neglecting the
decay in the present work.

The order of magnitude of the bounds derived in
Fig.~\ref{fig:scalbounds} can be quantitatively understood from the
comparison shown in Fig.~\ref{fig:pot} between the unparticle induced
potential function and that expected from a standard finite-range
interaction.  As seen in the figure, for any value of $d\lesssim 2$
the $W$ function is always larger than the corresponding one for a
$d=1$ force of range $\lambda=0.1\, R_\odot$.  Consequently, for any
value of $d\lesssim 2$ the bounds on the unparticle effects must be
comparable or stronger than the limits obtained in
Ref.~\cite{ourlrange} for a leptonic scalar, vector, or tensor,
long-range force with coupling $\kappa_S$, $\kappa_V$ and $\kappa_T$, 
respectively, and of range $\lambda=0.1 R_\odot $ with the identification
\begin{eqnarray}
\kappa_S&\rightarrow& \alpha_S 
\left(\frac{\Lambda_u}{M}\right)^{2(d_{UV}-1)} 
\frac{2}{\pi^{2d-1}}\frac{\Gamma[d+\frac{1}{2}]\Gamma[d-\frac{1}{2}]}
{\Gamma[2d]} \frac{1}{\left(R_\odot \Lambda_u\right)^{2(d-1)}}\leq
6.8\times 10^{-45}  
\label{eq:ksd} \, ,\\
\kappa_V&\rightarrow& \alpha_V 
\left(\frac{\Lambda_u}{M}\right)^{2(d_{UV}-1)} 
\frac{2}{\pi^{2d-1}}\frac{\Gamma[d+\frac{1}{2}]\Gamma[d-\frac{1}{2}]}
{\Gamma[2d]} \frac{1}{\left(R_\odot \Lambda_u\right)^{2(d-1)}}\leq
4.5\times 10^{-53} 
\label{eq:kvd} \, , \\
\kappa_T&\rightarrow& \alpha_T \frac{m_p} {\Lambda_u^2} B\,
\left(\frac{\Lambda_u}{M}\right)^{2(d_{UV}-1)} 
\frac{2}{\pi^{2d-1}}\frac{\Gamma[d+\frac{1}{2}]\Gamma[d-\frac{1}{2}]}
{\Gamma[2d]} \frac{1}{\left(R_\odot \Lambda_u\right)^{2(d-1)}} \leq
\frac{2\times 10^{-61}}{{\rm eV}}\; . 
\label{eq:ktd} 
\end{eqnarray}
For comparison we also plot in Fig.~\ref{fig:scalbounds} the bounds 
directly given by Eq.~\eqref{eq:ksd}. As seen the results from the full 
analysis are in reasonable quantitative agreement with the approximations
above.

From Eqs.~\eqref{eq:ksd}--~\eqref{eq:ktd} we can see that for a given
value of $\Lambda_u$ at the same scale dimension $d$, the bounds on
$M$ are $\sim 100$ times more stringent for vector unparticles. For
tensor unparticles the bounds on $M$ will be
$\sim 3/\sqrt{\Lambda_u/{\rm TeV}}$ times those  for the scalar
case. The bounds tend to get substantially relaxed as $d$ increases.
Consequently, , if the low energy scale is of the other 
$\Lambda_u\sim$ TeV,    
one cannot provide very significant constraints on vector and
tensor unparticles arising in a full conformal symmetry
-- for which $d>3,4$ respectively-- from their long-range effects on 
neutrino oscillation data.

\section{Discussion}
\label{sec:disc}

We have studied the effect of the long-range forces induced by the coupling of 
unparticle operators to fermions of the SM, which are exerted along the 
the neutrino trajectory from its production point inside the Sun 
to its detection at the Earth. 
We have used data from solar neutrino experiments and the newest KamLAND 
results to show that one can place stringent constraints on the parameters 
involved in the description of these new interactions.  

For a scalar unparticle, for instance, assuming the non-universality of 
the lepton couplings to unparticles to be of the order of a few per mil, 
{\it i.e.} $\alpha_s =10^{-3}$, and $d_{\rm UV}=3$, 
we find that the mass $M$ of the UV exchanged particle is constrained to be 
$M>8\times  10^{8}$ TeV ($M> 7\times 10^{9}$ TeV), for the scaling dimension $d=1.1$, if
the lower energy scale  is at $\Lambda_u =$ 1 TeV (10 TeV). 

In general the bounds derived here are much stronger than the reach at 
present and near future colliders. They are comparable with those 
imposed from cosmology and astrophysics and from the effect on 
the modification of the Newtonian law of gravity. For the sake of comparison 
we show in Table~\ref{tab1} our bounds together with those imposed by 
BBN~\cite{cosmo1}, and from the compilation in Table 2 of
Ref.~\cite{wyler} of the bounds from astrophysics and 5th force experiments.
In order to make this comparison we have assumed all $C^f_S$ to be
of order 1 but that some violation of universality is allowed at the
per mil level. 

\begin{table}
\begin{tabular}{|l|r|r|r|r|}
\hline
Scalar Unparticle & \multicolumn{4}{c|}{$M>$ (TeV)} \\
\hline
 $\Lambda_u$ & \multicolumn{2}{c|}{1 TeV} &
\multicolumn{2}{c|}{10 TeV} \\
\hline
d & 1.3 &  1.6 & 1.3 & 1.6 \\
\hline
This work & $1.3\times 10^{6}$ & $100$  & $9 \times 10^{6}$ & $500$  \\
\hline 
BBN &  $3.2\times 10^{5}$ & $5.8\times 10^3$ 
& $4.7\times 10^6$ & $1.2\times 10^6$\\
\hline
E\"{o}tv\"{o}s-type~ & 
$1.9\times 10^{6}$ & $3.5\times 10^{3}$
& $1.3 \times 10^{7}$ & $1.6 \times 10^{4}$ \\
\hline
Energy loss from stars~ & $800$ & $26$ & $5.5\times 10^3$ & $120$ \\
\hline
SN 1987A~ & $430$ & $55$ & $3\times 10^3$ & $260$ \\
\hline
\end{tabular}
\caption{Limits on $M$ from various sources testing for signals from
scalar unparticle couplings to fermions. Here $d_{\rm UV}=3$ was
used in deriving all bounds and $\alpha_s =10^{-3}$ was used for
our work.}
\label{tab1}
\end{table}
We can see that for low values of the scale $d$, the bounds
derived in this work are comparable to the most stringent ones, coming from 
scalar unparticle mediated 5th-force tests. However, unlike these last ones,
the bounds from the solar neutrino oscillation effects would apply even 
if the unparticles do not couple baryons. 

Conversely, the bounds we derived in this work can also be
converted into severe constraints on universality violation of the neutrino
couplings to unparticle operators with scales which can be accessible at 
future colliders.
For example, for $M=\Lambda_u=1$ TeV and $C_S^f={\cal O}(1)$, the bounds in 
Eq.~\eqref{eq:ksd} imply that
\begin{equation}
\frac{C^{\nu_e}_{S}-C^{\nu_a}_{S}}{C_S^{\nu_e}} \leq
4.5\times 10^{-38}- 1.5\times 10^{-9}   
\end{equation}
for $d=1.1$--$1.6$. 
Therefore in the absence of a model justification of why flavour
effects on the neutrino couplings to unparticle operators are more constrained
than on unparticle couplings for other fermions, this analysis implies that most
likely the fermion couplings to unparticles which can be tested at colliders
will be flavour blind.

\vspace{-0.3cm}
\begin{acknowledgments} 
\vspace{-0.3cm}
One of us (R.Z.F) is grateful for the hospitality of the
Departament d'Estructura i Constituents de la Mat\`eria, 
Universitat de Barcelona during part of this work.
This work is supported by National Science Foundation
grant PHY-0354776, by Spanish Grants  FPA-2004-00996, 
by Funda\c{c}\~ao de Amparo \`a Pesquisa do Estado 
de S\~ao Paulo (FAPESP) and by Conselho Nacional de Ci\^encia e
Tecnologia (CNPq). 
\end{acknowledgments}


\end{document}